\documentclass[prb,aps,groupedaddress,floatfix,twocolumn,superscriptaddress,showpacs]{revtex4}
\usepackage{amsmath,amssymb,epsfig,bm,pifont}
\usepackage[linkcolor=black,citecolor=black,urlcolor=blue,colorlinks=true,
linktocpage=true]{hyperref}
\usepackage{graphicx}
\usepackage{epstopdf}
\usepackage{epsfig}
\usepackage{bm}
\usepackage{tabularx}

\newcommand{\beq}{\begin{equation}}
\newcommand{\eeq}{\end{equation}}
\newcommand{\bea}{\begin{eqnarray}}
\newcommand{\eea}{\end{eqnarray}}
\newcommand{\tr}{\mathrm{tr}}

\newcommand{\bd}{\boldsymbol}

\newcommand{\mrm}{\mathrm}

\newcommand{\erf}{\mathrm{erf}}
\newcommand{\veps}{\varepsilon}

\begin{document}

\title{Entanglement of Free Fermions and Bosons at Finite Temperature}

\author{Joaqu\'{\i}n E. Drut}
\email{drut@email.unc.edu}
\affiliation{Department of Physics and Astronomy, University of North Carolina,
Chapel Hill, North Carolina, 27599-3255, USA}

\author{William J. Porter}
\email{wjporter@live.unc.edu}
\affiliation{Department of Physics and Astronomy, University of North Carolina,
Chapel Hill, North Carolina, 27599-3255, USA}

\begin{abstract}
We generalize techniques previously used to compute ground-state properties of one-dimensional noninteracting quantum gases to obtain exact results at finite temperature.  We compute the order-$n$ R\'enyi entanglement entropy to all orders in the fugacity in one, two, and three spatial dimensions.  In all spatial dimensions, we provide closed-form expressions for its virial expansion up to next-to-leading order.  In all of our results, we find explicit volume scaling in the high-temperature limit.
\end{abstract}

\date{\today}
\pacs{03.65.Ud, 05.30.Fk, 03.67.Mn}
\maketitle

\section{Introduction} 

The exciting prospect of quantum computation for condensed matter, nuclear, and high-energy physics
has brought about considerable activity at the interface between those areas and that of quantum information~\cite{ReviewQI}.
This interest has been fueled even further by the discovery of direct connections between topological quantum
phase transitions and the concept of entanglement entropy~\cite{EEandQPT}. More recently, the realization of topological
insulators in condensed matter systems has brought all of the above together and is at the moment one of the most active areas
of research in physics (see e.g.~\cite{QCTI}).

In the last few years, substantial work has been carried out to rigorously characterize the scaling of the R\'enyi and
von Neumann entanglement entropies of noninteracting systems in the ground state~\cite{ScalingGS}. A number of exact results 
and inequalities have been established for a variety of systems (e.g., in external traps,
in varying dimensions, etc.) in the ground state~\cite{MiscExactGases2009,MiscExactGases2012,MiscExactGases2013,
MiscExactGases2014,MiscExactGases2015,MiscExactSpinChains2012,MiscExactSpinChains2014,MiscExactSpinChainsGapped}. In contrast, the behavior of the R\'enyi 
entanglement entropy for such systems at finite temperature has been much less explored~\cite{FiniteTemp}.

To help fill that gap, we generalize a technique previously applied with great success to degenerate noninteracting gases~\cite{OverlapMatrix2011,OverlapMatrix2012}. 
Using this method, we perform calculations of the R\'enyi entanglement entropies $S_n^{}$ to all orders in the fugacity $z=e^{\beta \mu}$, where $\beta$ is 
the inverse temperature and $\mu$ the chemical potential.
In addition, we present closed-form expressions of the contributions to $S_n^{}$ at
leading and next-to leading order in $z$ around $z=0$, which are the first terms in the 
virial expansion; we compute those contributions for both bosons and fermions. This expansion is interesting
because, even though it is only valid in a dilute regime (which implies high temperature in the scale set by the Fermi energy),
it is non-perturbative in character, and it is therefore useful as a benchmark for other approaches toward 
calculating what has proven to be a challenging observable. Those approaches include, in particular, quantum Monte Carlo methods, 
of which a variety of flavors have been recently put forward~\cite{Buividovich,Melko,Humeniuk,McMinis,Grover,Broecker,WangTroyer,Luitz,Assaad,HMCEE}.

It should be noted that we do not expect to see in our results any kind of peculiar behavior associated with phase transitions 
(such as anomalous area-law scaling). However, we emphasize that the results are exact and for this reason they are both 
useful and interesting. 

\section{Formalism and results for fermions}

\subsection{Basics}

We set the stage for our calculation of noninteracting fermions in this section.
We focus on non-relativistic $d$-dimensional systems, such that the dispersion relation is $\veps(\bd{p}) = \bd{p}^2/2m$, taking $m = 1$ in what follows. The grand-canonical partition function is
\beq
\mathcal{Z} = \det\left(\openone + z\,\mathcal{U}\right)
\eeq
where $\mathcal{U}_{\bd{p}\bd{p}'} = \delta^{(d)}(\bd{p} - \bd{p}') \exp(-\beta \veps(\bd{p}))$,
$z=\exp(\beta\mu)$ is the fugacity, $\beta$ is the inverse temperature and $\mu$ is the chemical potential.  The single-particle Green's function is
\beq
\mathcal{G} = \frac{z\,\mathcal{U}}{\openone + z\,\mathcal{U}},
\eeq
which is diagonal in momentum space:
\beq
\mathcal{G}_{\bd{p}\bd{p}'} = \frac{z e^{-\beta \veps(\bd{p})}}{1+z e^{-\beta \veps(\bd{p})}}\;\delta^{(d)}(\bd{p} - \bd{p}'),
\eeq
while in coordinate space (in an infinite volume) it takes the form
\beq
\mathcal{G}_{\bd{x}\bd{x}'} = \int_{\mathbb R^d}\frac{d^dp}{(2\pi)^d}e^{i\bd{p}\cdot(\bd{x}-\bd{x}')}
\frac{z e^{-\beta \veps(\bd{p})}}{1+z e^{-\beta \veps(\bd{p})}}.
\eeq

To characterize the behavior of the entanglement entropy in the low-$z$ regime, we need $\mathcal G$ in the virial expansion,
which we write as
\beq
\label{Eq:GFermionsVirial}
\mathcal{G}_{\bd{x}\bd{x}'} = \frac{z}{\lambda^d_T} \sum_{j=0}^{\infty} \mathcal{A}_{j,\bd{x}\bd{x}'}z^j,
\eeq
where the thermal wavelength $\lambda^{}_T = \sqrt{2\pi\beta}$ and the expansion coefficients are functions
given by
\beq
\mathcal{A}^{}_{j,\bd{x}\bd{x}'}=\frac{(-1)^j}{(j+1)^{d/2}}\exp\left(-\frac{\pi |\bd{x}-\bd{x}'|^2}{\lambda^2_T (j+1)}\right).
\eeq
These functions will be essential in the virial expansion of the entanglement entropy, which we discuss next.

\subsection{Entanglement entropy}

The R\'enyi entanglement entropy $S_n$ for a subregion $A$ is defined for positive integers $n$ from the system's reduced density matrix $\hat \rho_A$ as 
\beq
(1-n) S_n = \ln\tr\, \hat \rho^{n}_{A}.
\eeq
By way of an auxiliary field decomposition, Grover showed that $S_n$ may be computed as an expectation value taken over multiple copies (replicas) of a Hubbard-Stratonovich (HS) field (see Ref.~\cite{Grover}).  Writing the reduced density operator as a superposition of operators corresponding to free particles in an HS field $\sigma$,
\beq
\hat \rho^{}_A = \int \mathcal D \sigma \; P_\sigma \,\hat \rho^{}_{A,\sigma},
\eeq
with the normalized probability measure $P_\sigma$ determined by the underlying dynamics~\cite{}, an explicit form for the operators $\hat \rho_{A,\sigma}$ may be derived by requiring them to reproduce the one-body physics.  In particular, Grover showed
\beq
\hat \rho^{}_{A,\sigma} = C_{A,\sigma}\exp\left(-\bd{\hat{c}}^{\dagger}\log\left(\mathcal{G}_{A,\sigma}^{-1} - \mrm{Id}_A\right)\bd{\hat{c}}\right),
\eeq
where $\bd{\hat{c}}^{(\dagger)}$ are the fermionic annihilation (respectively, creation) operators, and
\beq
C^{}_{A,\sigma} = \det\left(\mrm{Id}_A - \mathcal{G}_{A,\sigma}\right)
\eeq
preserves the unit normalization. The identity operator (continuous or discrete) supported in the region $A$ has been denoted 
by $\mrm{Id}_A$, and $\mathcal{G}_{A,\sigma}$ is the single-particle restricted Green's function in background field $\sigma$.

Using this form for the reduced density operator, it follows that the R\'enyi entropy can be written in path integral form as
\beq
\label{Eq:SnMn}
(1-n)S^{}_n = \ln \int \mathcal D \bd{\sigma} P[\bd{\sigma}]\det \mathcal{M}_n[\bd{\sigma}]
\eeq
with $\bd{\sigma} = (\sigma_1,\dots,\sigma_n)$, for normalized measure
\beq
\mathcal D \bd{\sigma} P[\bd{\sigma}] = \mathcal D\sigma_1\dots\mathcal D\sigma_n\;P_{\sigma_1}\dots P_{\sigma_n},
\eeq
and with
\begin{align}
\mathcal{M}^{}_n[\bd{\sigma}] &= \prod_{k=1}^{n}\left(\mrm{Id}_A-\mathcal{G}_{A,\sigma_k}\right)\times \nonumber \\
&\;\;\;\left[\mrm{Id}_A + \prod_{k=1}^{n}\frac{\mathcal{G}_{A,\sigma_k}}{\mrm{Id}_A-\mathcal{G}_{A,\sigma_k}}\right].
\end{align}

Despite its amazingly general form, this expression has been found to be ill-suited for direct numerical calculations for systems with interactions of any appreciable strength, even in the case where $n=2$ which requires no matrix inversion~\cite{Assaad}.  In spite of its complexity, numerous complementary approaches have been devised to compute the R\'enyi entropies for interacting systems~\cite{HMCEE,Buividovich,Melko,Humeniuk,McMinis,Broecker,WangTroyer,Luitz,Assaad}.

For the noninteracting case, the measure is trivial, and the entropy may be recast as
\beq
\label{Eq:SnMnFree}
(1-n) S_n = \ln\det \mathcal M^{}_n
\eeq
where $\mathcal{M}_n$ is given 
\beq
\mathcal M^{}_n = (\mrm{Id}_{A} - \mathcal G_{A})^n + \mathcal{G}_{A}^n,
\eeq
and multiplications prescribed above, in the case of continuous quantum numbers, are to be understood as
\beq
[\mathcal{O}^{}_1\mathcal{O}^{}_2]_{\bd{x}\bd{x}'} = \int_{A}d^dy\;\mathcal{O}_{1,\bd{x}\bd{y}}\mathcal{O}_{2,\bd{y}\bd{x}'}
\eeq
given operators $\mathcal{O}_1$ and $\mathcal{O}_1$ supported for $\bd{x},\bd{x}'\in A$.

The above results for noninteracting systems, or in general for quadratic Hamiltonians, were first derived by Peschel and
colleagues in Refs.~\cite{Peschel}.

\subsection{Entanglement entropy from overlap matrices}

Inspired by Refs.~\cite{OverlapMatrix2011,OverlapMatrix2012}, we place the system in a finite box of volume $V = L^d$ and implement periodic boundary conditions.  We define the Fredholm determinant $D_{A}(\lambda)$ in terms of the restricted finite-temperature correlator $\mathcal{G}_A$ and the identity operator on our subregion $\mrm{Id}_A$ via
\beq
D_{A}(\lambda) = \det\left(\lambda\,\mrm{Id}_A  - \mathcal{G}_A\right),
\eeq
and as previously done in Ref.~\cite{ToeplitzJinKorepin}, we write the $n$-th order R\'enyi entropy as a contour integral
\beq
S_n = \oint_{\Gamma}\frac{d\lambda}{2\pi i}\;f_n(\lambda)\frac{d}{d\lambda}\ln D_{A}(\lambda)
\eeq
where
\beq
f_{n}(\lambda) = \frac{1}{1-n}\ln\left[(1-\lambda)^n + \lambda^n\right],
\eeq
and the integration contour $\Gamma$ encloses the segment $[0,1]\subset \mathbb R$ of the complex-$\lambda$ plane.  From this expression, we write
\beq
\ln \det\left(\lambda\,\mrm{Id}_A  - \mathcal{G}_A\right) = \omega(V) \ln \lambda + \ln \det\left(\mrm{Id}_A  -\lambda^{-1}\mathcal{G}_A\right) 
\eeq
for some (infinite) constant $\omega$ that depends on the volume of space.  Because the function $f_n$ vanishes at $\lambda=0$ for $n\ge 2$, this term makes no contribution to the integral, and we may safely neglect it in derivations that follow and in the infinite-volume limit taken later.  Expanding the Fredholm determinant on the right in powers of $\lambda^{-1}$, we have
\beq
\ln \det\left(\mrm{Id}_A  -\lambda^{-1}\mathcal{G}_A\right) = -\sum_{s=1}^{\infty}\frac{\tr\,\mathcal{G}^s_A}{\lambda^s s},
\eeq
 and writing out the trace in $d$ spatial dimensions, we obtain
 \beq
 \tr\,\mathcal{G}^s_A = \sum_{\bd{k_1}\in \mathbb Z^d}\dots\sum_{\bd{k_s}\in \mathbb Z^d}A_{\bd{k_1}\bd{k_2}} A_{\bd{k_2}\bd{k_3}}\dots A_{\bd{k_{s}}\bd{k_1}} = \tr\,A^s
 \eeq
 for each positive integer $s$ where the (infinite) matrix $A$ is comprised of elements
 \beq
 A_{\bd{k}\bd{k'}} =\frac{1}{L^d} \frac{z e^{-\pi \bd{k}^2 (\lambda_T/L)^2}}{1+z e^{-\pi \bd{k}^2 (\lambda_T/L)^2}}\int_{A}d^dx\;e^{-2\pi i \bd{x}\cdot(\bd{k}-\bd{k'})/L}.
\eeq
We note that this matrix is identical to that used in Refs.~\cite{OverlapMatrix2011,OverlapMatrix2012} save two crucial differences:  The domain of the indices is not bounded as a result of all momentum states having nontrivial occupation at finite temperature; this noncompact support is thankfully accompanied by the additional Fermi-Dirac prefactor providing a point of connection between our generalization and this method's ground-state formulation 
(again, see Refs.~\cite{OverlapMatrix2011,OverlapMatrix2012}).  In the case of continuous indices, we find
\beq
S_n = \int_{\mathbb{R}^d}d^d k\;f_n(a(\bd{k}))
\eeq
where the values $a(\bd{k})$ are the spectrum of the continuum limit of $A$, and we have indexed them by momenta.

In order to perform our numerical calculations for arbitrary fugacity, we diagonalize the matrix $A$ defined above at fixed 
\beq
\xi \equiv \lambda_T/L,
\eeq
for some finite momentum cutoff $\Lambda$.  Following that, we formally construct a (diagonal) matrix $M_n^{(\Lambda)}$ with elements $f_n(a(\bd{k}))\delta_{\bd{k}\bd{k'}}$ for values of the $d$-dimensional integer vectors satisfying $|\bd{k}|,|\bd{k'}|\le \Lambda$ and perform the trace corresponding to the continuum limit via the limit of infinite cutoff and consequently the limit in which the matrix becomes infinite in linear extent $N_{\Lambda}$.  That is, we compute the entropy in a finite volume via the limit
\beq
S_n^{(\Lambda)}(\xi) \equiv \frac{1}{N_\Lambda}\tr\,M_n^{(\Lambda)}\xrightarrow[]{N_{\Lambda}\to\infty}\int_{\mathbb{R}^d}d^d k\;M_n(\bd{k},\bd{k}).
\eeq
In addition to the continuum limit, we perform the infinite volume limit by taking $\xi$ to zero while fixing the region size $L_A$ in units of the thermal wavelength, thus maintaining that we are in the dilute (high-temperature) regime.

From the above definitions, we note that for $\xi\ll 1$, the matrix elements of $A$ become
\beq
 A_{\bd{k}\bd{k'}} = \frac{(\xi\theta)^d}{(2\pi)^{d/2}} \frac{z}{1+z} + \mathcal{O}(\xi^{d+2}),
\eeq
where we have written
\beq
\theta =\frac{L_A}{\lambda_T}\sqrt{2\pi};
\eeq
that is, they become momentum-independent to leading order as we take the continuum limit, so that in the this limit, $A$ is a constant matrix.  Working at leading order, we find that for fermions this gross degeneracy implies that the entropy takes a particularly simple form.  Independent of the form of the cutoff, we find that for finite $\Lambda$ the above trace gives
\beq
(\xi \theta)^{-d}S_n^{(\Lambda)}(\xi)\xrightarrow[]{\xi\to 0}\frac{1}{(2\pi)^{d/2}}\frac{n}{n-1}\frac{z}{1+z},
\eeq
a result independent of $\Lambda$.  

\subsection{Virial expansion of the entropy for fermions}

Starting from Eq.~(\ref{Eq:SnMn}), where the Green's function is an explicit function of the fugacity $z$, we expand to 
leading and next-to-leading order in $z$. In the interacting case, this involves expanding both the determinant present in the probability
measure $P$, as well as the pure entanglement contribution coming from $\det \mathcal{M}_n$. In the noninteracting case,
however, $P$ is just a constant and therefore plays no role in the calculation, i.e. our starting point is actually Eq.~(\ref{Eq:SnMnFree}). 
Thus, expanding that equation in powers of $z$, we obtain
\bea
(1-n)S_n^{} = &&\left. \tr \ln \mathcal{M}_n\right |_{z=0} \\ \nonumber
&& \!\!\!\!\!\!\!\!\!\!\!\!\!\!\!\!\!\!\!\!\!\!\!\! + z \left. \tr \left[\mathcal{M}^{-1}_n \frac{\partial \mathcal{M}^{}_n}{\partial z}\right]  \right |_{z=0} \\ \nonumber
&& \!\!\!\!\!\!\!\!\!\!\!\!\!\!\!\!\!\!\!\!\!\!\!\! + \frac{z^2}{2!}\left. \left\{  \tr \left[\mathcal{M}^{-1}_n \left(\frac{\partial^2 \mathcal{M}^{}_n}{\partial z^2}\right) \!-\! 
\mathcal{M}^{-2}_n \left(\frac{\partial \mathcal{M}^{}_n}{\partial z}\right)^2 \right] \right\}  \right |_{z=0} \\ \nonumber + O(z^3).
\eea

We continue by analyzing the $z$ dependence of $\mathcal{M}_n$.
At $z = 0$, the single-particle Green's function vanishes, which implies that the first term in the above expansion vanishes as well.
From this, we find immediately that the limiting form of the operator $\mathcal{M}_n$, as well as its inverse (where it exists), is the identity,
i.e.
\beq
\left. \mathcal{M}_n\right |_{z=0} = \left. \mathcal{M}^{-1}_n\right |_{z=0} = \mrm{Id}_{A}.
\eeq
 
Therefore,
\begin{align}
(1-n)S_n^{}  &= z\,\tr\left[\frac{\partial \mathcal{M}_n}{\partial z}\right]\Bigg\vert_{z=0} \\ 
&+ \frac{z^2}{2!}\,\tr\left[\frac{\partial^2 \mathcal{M}_n}{\partial z^2}-\left(\frac{\partial \mathcal{M}_n}{\partial z}\right)^2\right]\Bigg\vert_{z=0} + O(z^3). \nonumber
\end{align}

Further, we have for $n\ge 2$
\beq
\frac{\partial \mathcal{M}^{}_n}{\partial z}\Bigg\vert_{z=0} = -n \frac{\partial \mathcal{G}_{A}}{\partial z}\Bigg\vert_{z=0}.
\eeq
The peculiar structure of the operator under consideration, together with the behavior of the Green's function at $z = 0$,
necessitates separating the second order contribution into cases $n =2$ and $n > 2$.  For $n = 2$, we find
\beq
\frac{\partial^2 \mathcal{M}^{}_2}{\partial z^2}\Bigg\vert_{z=0} =\left[4  \left(\frac{\partial \mathcal{G}_{A}}{\partial z}\right)^2 - 2 \frac{\partial^2 \mathcal{G}_{A}}{\partial z^2}\right]\Bigg\vert_{z=0},
\eeq
with the general expression for $n > 2$ being different:
\beq
\frac{\partial^2 \mathcal{M}^{}_n}{\partial z^2}\Bigg\vert_{z=0} =\left[n(n-1)  \left(\frac{\partial \mathcal{G}_{A}}{\partial z}\right)^2 - n \frac{\partial^2 \mathcal{G}_{A}}{\partial z^2}\right]\Bigg\vert_{z=0}.
\eeq
Therefore,
\beq
\left[\frac{\partial^2 \mathcal{M}_2}{\partial z^2}-\left(\frac{\partial \mathcal{M}_2}{\partial z}\right)^2\right]\Bigg\vert_{z=0} =
- 2 \frac{\partial^2 \mathcal{G}_{A}}{\partial z^2}\Bigg\vert_{z=0}.
\eeq
and
\beq
\left[\frac{\partial^2 \mathcal{M}_n}{\partial z^2}-\left(\frac{\partial \mathcal{M}_n}{\partial z}\right)^2\right]\Bigg\vert_{z=0} =
-n \left[\left(\frac{\partial \mathcal{G}_{A}}{\partial z}\right)^2 + \frac{\partial^2 \mathcal{G}_{A}}{\partial z^2}
\right]\Bigg\vert_{z=0}
\eeq

From the virial expansion of the free fermionic Green's function given in Eq.~(\ref{Eq:GFermionsVirial}), we observe not only
\beq
\frac{\partial \mathcal{G}_{\bd{x}\bd{x}'}}{\partial z}\Bigg\vert_{z=0} =   \lambda^{-d}_T\mathcal{A}_{0,\bd{x}\bd{x}'}
\eeq
but also
\beq
\frac{\partial^2 \mathcal{G}_{\bd{x}\bd{x}'}}{\partial z^2}\Bigg\vert_{z=0} =  2\lambda^{-d}_T\mathcal{A}_{1,\bd{x}\bd{x}'}.
\eeq
The restricted form of $\mathcal{G}_{\bd{x}\bd{x}'}$, therefore, translates into restricted forms of the 
$\mathcal{A}_{j,\bd{x}\bd{x}'}$ coefficients.

\begin{figure}[h!]
\includegraphics[width=1.0\columnwidth]{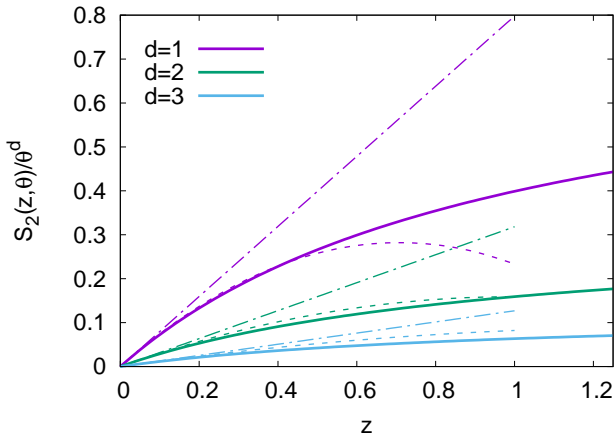}
\caption{\label{Fig:S2FugacityFermions}(color online) Analytic results to all orders in $z$ (solid lines) along with first- (dot-dashed) and second-order (dashed) virial expansions for the second R\'enyi entropy $S_2(z,\theta)$ for fermions versus fugacity, shown for dimensions $d=1$, $2$, and $3$ (from top to bottom). Note that this case ($n=2$) has no dependence on $\theta$ at the orders shown.
}
\end{figure}

We evaluate the required traces according to
\beq
\tr\,\mathcal{O}_A = \int_A d^d y \;\mathcal{O}_{A,\bd{y}\bd{y}}
\eeq
for the $d$-dimensional case, and we have chosen the region $A$ to be the rectangular prism defined by the vectors $\bd{x_1}$ and $\bd{x_2}$ via $ A = [x_{1(1)},x_{2(1)}]\times\dots\times[x_{1(d)},x_{2(d)}]\subseteq \mathbb{R}^d$
Putting the above contributions together up to second order in $z$, we find that the second R\'enyi entropy is
\beq
\label{Eq:S2z2}
S_2(z,\theta) = \left(2\, \tr \,\mathcal{A}^{}_{0,A}\right)\, z + \left(2\, \tr\, \mathcal{A}^{}_{1,A}\right)\, z^2 + O(z^3),
\eeq
while for $n>2$ R\'enyi entropies we obtain
\begin{align}
\label{Eq:Snz2}
S_n(z,\theta) &= \frac{n}{n-1}\left(\tr \,\mathcal{A}^{}_{0,A}\right)\, z \\
&\;\;\;+ \frac{n}{2(n-1)}\left(\tr \,\mathcal{A}^{2}_{0,A} + 2\, \tr\, \mathcal{A}^{}_{1,A}\right)\, z^2 + O(z^3) \nonumber.
\end{align}
For the coefficients, we obtain the following closed-form expressions:
\beq
\tr\, \mathcal{A}^{}_{0,A} =(2\pi)^{-d/2} \theta^{}_1\theta^{}_2\dots\theta^{}_d
\eeq
\beq
\tr \,\mathcal{A}^{}_{1,A} = -\frac{1}{2^{d/2}}\tr \,\mathcal{A}_{0,A} = -\frac{1}{2^d\pi^{d/2}} \theta^{}_1\theta^{}_2\dots\theta^{}_d
\eeq
\beq
\tr \,\mathcal{A}^{2}_{0,A} = \prod_{k=1}^{d}\Bigg[\frac{1}{2\sqrt{\pi}}\,\theta^{}_k\erf\, \theta^{}_k + \frac{1}{2\pi}\left(e^{-\theta^{2}_k}-1\right)\Bigg].
\eeq

In each case above, we have introduced dimensionless length variables $\theta_k$ defined for each $k$ by
\beq
\theta^{}_k= \frac{\sqrt{2\pi}}{\lambda_T}\left(x_{2(k)}-x_{1(k)}\right).
\eeq

Equations (\ref{Eq:S2z2}) and (\ref{Eq:Snz2}) are our main result for the virial expansion for fermions. Plots of the first of these results are given in Fig.~\ref{Fig:S2FugacityFermions}, where we display the second R\'enyi entropy $S_2(z,\theta)$ for dimensions $d = 1$, $2$, and $3$.
We note that for the second entropy, both virial coefficients display explicit volume scaling so that to quadratic order in the fugacity, 
$S_2(z,\theta)/\theta^d$ is a function of the fugacity and spatial dimension alone. 
\begin{figure}[h]
\includegraphics[width=1.0\columnwidth]{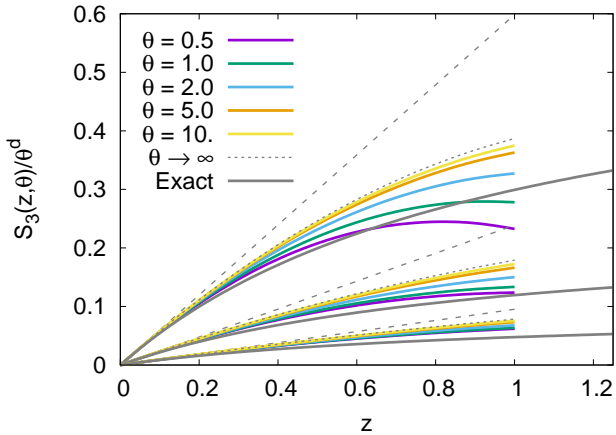}
\caption{\label{Fig:S3FugacityFermions}(color online) 
Third R\'enyi entropy $S_3^{}$ of noninteracting fermions as a function of the fugacity and for several values
of the dimensionless region size $\theta$ and for dimensions 1, 2 and 3 (top to bottom groups). The dashed lines show,
for each dimension, the leading order results, which are proportional to $\theta^d$, colored lines represent the second-order results for various subsystem sizes, and solid extended lines show the result of analytic calculations to all orders in the fugacity.
}
\end{figure}
In Fig.~\ref{Fig:S3FugacityFermions}, we show the $n=3$ case for several values of the dimensionless region size $\theta$ and 
for dimensions 1, 2 and 3. In contrast to the $n=2$ case, $n>2$ does have non-trivial volume effects at second order in $z$.

\section{Formalism and results for bosons}

\subsection{Basics}

Continuing with our focus on non-relativistic systems, the partition function for noninteracting bosons is
\beq
\mathcal{Z} = \left[\det\left(\openone - z\,\mathcal{U}\right)\right]^{-1},
\eeq
such that the single-particle Green's function is
\beq
\mathcal{G} = \frac{z\,\mathcal{U}}{\openone - z\,\mathcal{U}},
\eeq
which takes the momentum-space form
\beq
\mathcal{G}_{\bd{p}\bd{p}'} = \frac{z e^{-\beta \veps(\bd{p})}}{1-z e^{-\beta \veps(\bd{p})}}\;\delta^{(d)}(\bd{p} - \bd{p}'),
\eeq
and the coordinate-space form $\mathcal{G}_{\bd{x}\bd{x}'}$ can be obtained by Fourier transforming, as shown above in the fermion case.
With this as a starting point, we may formally expand in powers of $z$:
\beq
\mathcal{G}_{\bd{x}\bd{x}'} = \frac{z}{\lambda^d_T} \sum_{j=0}^{\infty} \mathcal{B}_{j,\bd{x}\bd{x}'}z^j
\eeq
where, not surprisingly, we find
\beq
\mathcal{B}_{j,\bd{x}\bd{x}'} = (-1)^{j} \mathcal{A}_{j,\bd{x}\bd{x}'}.
\eeq
%

\subsection{Entanglement entropy}

With the replacement of the bosonic Green's function defined above for the fermionic equivalent used earlier, the auxiliary-field decomposition of the reduced density matrix becomes
\beq
\hat \rho_{A,\sigma} = B_{A,\sigma}\exp\left(-\bd{\hat{b}}^{\dagger}\log\left(\mathcal{G}_{A}^{-1} + \mrm{Id}_A\right)\bd{\hat{b}}\right)
\eeq
where
\beq
B_{A,\sigma} = \left[\det\left(\mrm{Id}_A + \mathcal{G}_{A}\right)\right]^{-1}
\eeq
and we have introduced the bosonic annihilation (creation) operators $\bd{\hat{b}}^{(\dagger)}$.

From this expression for the reduced density operator, the R\'enyi entropy for bosons can be expressed in path integral form as
\beq
(1-n)S_n = \ln \int \mathcal D \bd{\sigma} P[\bd{\sigma}]\det \mathcal{N}_n[\bd{\sigma}]
\eeq
with
\begin{align}
\mathcal{N}_n[\bd{\sigma}] &= \prod_{k=1}^{n}\left(\mrm{Id}_A+\mathcal{G}_{A,\sigma_k}\right)^{-1}\times \nonumber \\
&\;\;\;\left[\mrm{Id}_A - \prod_{k=1}^{n}\frac{\mathcal{G}_{A,\sigma_k}}{\mrm{Id}_A+\mathcal{G}_{A,\sigma_k}}\right]^{-1}
\end{align}
and where the probability $P[\bd \sigma]$ is now generated by the product measure of the single-field densities coming from the underlying dynamics of the bosonic degrees of freedom.

Interestingly, the above expressions extend Grover's Monte Carlo method to interacting bosons. However, stability of the
latter requires a repulsive interaction, which would give a sign problem. Therefore, the above is not (at the moment)
of practical interest for Monte Carlo methods. It does, however, provide a starting point for formal analyses.
In particular, in the noninteracting case the entropy becomes
\beq
(1-n) S_n = - \ln\det\mathcal{N}_n
\eeq
where
\beq
\mathcal{N}_n =  (\mrm{Id}_{A} + \mathcal G_{A})^n - \mathcal{G}_{A}^n.
\eeq
%

\subsection{Bosonic entanglement entropy from overlap matrices}
In close analogy to the fermionic case, we again place the system in a finite box and impose periodic boundary conditions.  We define the Fredholm determinant $D_{A}(\lambda)$ in terms of the restricted bosonic correlator as
\beq
D_{A}(\lambda) = \det\left(\lambda\,\mrm{Id}_A  - \mathcal{G}_A\right),
\eeq
and write the $n$-th order R\'enyi entropy as a contour integral
\beq
S_n = \oint_{\Sigma}\frac{d\lambda}{2\pi i}\;g_n(\lambda)\frac{d}{d\lambda}\ln D_{A}(\lambda)
\eeq
where now
\beq
g_{n}(\lambda) = \frac{1}{1-n}\ln\left[\frac{1}{(1+\lambda)^n - \lambda^n}\right],
\eeq
and the integration contour $\Sigma$ is defined to contain the segment $[0,\infty)\subset \mathbb R$ of the complex-$\lambda$ plane. 

Again, writing out the trace in $d$ spatial dimensions, we obtain an expression
 \beq
 \tr\,\mathcal{G}^s_A = \sum_{\bd{k_1}\in \mathbb Z^d}\dots\sum_{\bd{k_s}\in \mathbb Z^d}B_{\bd{k_1}\bd{k_2}} B_{\bd{k_2}\bd{k_3}}\dots B_{\bd{k_{s}}\bd{k_1}} = \tr\,B^s
 \eeq
 for each positive integer $s$ where the (again infinite) matrix $B$ differs from the matrix $A$ in an obvious way:
 \beq
 B_{\bd{k}\bd{k'}} =\frac{1}{L^d} \frac{z e^{-\pi \bd{k}^2\xi^2}}{1-z e^{-\pi \bd{k}^2\xi^2}}\int_{A}d^dx\;e^{-2\pi i \bd{x}\cdot(\bd{k}-\bd{k'})/L}.
\eeq
Evaluating the limit as before, we obtain a continuum limit valid to all orders in the fugacity:
\beq
(\xi \theta)^{-d}S_n^{(\Lambda)}(\xi)\xrightarrow[]{\xi\to 0}\frac{1}{(2\pi)^{d/2}}\frac{n}{n-1}\frac{z}{1-z}.
\eeq
%

\subsection{Virial expansion of the entropy for bosons}
Again, we find that a zero fugacity, the single-particle Green's function vanishes. Hence, once again in this limit, $\mathcal{N}_n$ and its inverse are the identity.  Therefore, the virial expansion of $S_n^{}$ in the bosonic case simplifies to
\begin{align}
(n-1)S_n^{}  &= z\,\tr\left[\frac{\partial \mathcal{N}_n}{\partial z}\right]\Bigg\vert_{z=0} \\ 
&+ \frac{z^2}{2!}\,\tr\left[\frac{\partial^2 \mathcal{N}_n}{\partial z^2}-\left(\frac{\partial \mathcal{N}_n}{\partial z}\right)^2\right]\Bigg\vert_{z=0} + O(z^3). \nonumber
\end{align}

We easily obtain for orders $n\ge 2$
\beq
\frac{\partial \mathcal{N}_n}{\partial z}\Bigg\vert_{z=0} = n \frac{\partial \mathcal{G}_{A}}{\partial z}\Bigg\vert_{z=0}.
\eeq
Similary operator structure requires we again treat separately the second order contribution into cases of $n =2$ and $n > 2$.  For $n = 2$, we get
\beq
\frac{\partial^2 \mathcal{N}_2}{\partial z^2}\Bigg\vert_{z=0} =2 \frac{\partial^2 \mathcal{G}_{A}}{\partial z^2}\Bigg\vert_{z=0}
\eeq
and with the expression for $n > 2$ being
\beq
\frac{\partial^2 \mathcal{N}_n}{\partial z^2}\Bigg\vert_{z=0} =\left[n(n-1)  \left(\frac{\partial \mathcal{G}_{A}}{\partial z}\right)^2 + n \frac{\partial^2 \mathcal{G}_{A}}{\partial z^2}\right]\Bigg\vert_{z=0}.
\eeq

From the virial expansion given above for the bosonic correlator, we find that for the second order entropy the expansion becomes
\begin{align}
\label{Eq:S2Bosons}
S_2(z,\theta) &= \left(2\, \tr \,\mathcal{B}^{}_{0,A}\right)\, z \\
&\;\;\;+ \left(2\, \tr\, \mathcal{B}^{}_{1,A} - 2\,\tr\, \mathcal{B}^{2}_{0,A}\right)\, z^2 + O(z^3) \nonumber \\
&= \left(2\, \tr \,\mathcal{A}^{}_{0,A}\right)\, z \\
&\;\;\;- \left(2\, \tr\, \mathcal{A}^{}_{1,A} + 2\,\tr\, \mathcal{A}^{2}_{0,A}\right)\, z^2 + O(z^3), \nonumber
\end{align}
and for the entropies of order $n>2$, we have
\begin{align}
\label{Eq:SnBosons}
S_n(z,\theta) &= \frac{n}{n-1}\left(\tr \,\mathcal{B}^{}_{0,A}\right)\, z \\
&\;\;\;+ \frac{n}{2(n-1)}\left(2\, \tr\, \mathcal{B}^{}_{1,A} - \tr \,\mathcal{B}^{2}_{0,A}\right)\, z^2 + O(z^3) \nonumber \\
&= \frac{n}{n-1}\left(\tr \,\mathcal{A}^{}_{0,A}\right)\, z \\
&\;\;\;- \frac{n}{2(n-1)}\left(\tr \,\mathcal{A}^{2}_{0,A} + 2\, \tr\, \mathcal{A}^{}_{1,A} \right)\, z^2 + O(z^3).\nonumber
\end{align}
\begin{figure}[t]
\includegraphics[width=1.0\columnwidth]{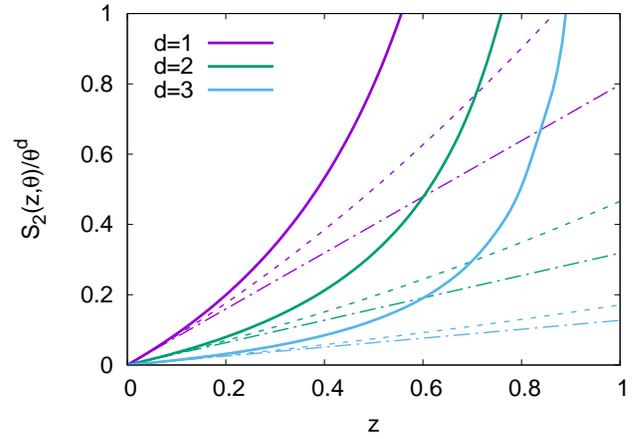}
\caption{\label{Fig:S2FugacityBosons}(color online) 
Exact results (solid) along with first- (dot-dashed) and second-order (dashed) virial expansions of the second R\'enyi entropy $S_2(z,\theta)$ for 
bosons shown (from top to bottom) for dimensions $d = 1$, $2$, and $3$ versus fugacity $z$.
}
\end{figure}

Equations~(\ref{Eq:S2Bosons}) and~(\ref{Eq:SnBosons}) are our main result for the virial expansion for bosons. As could be anticipated,
the leading-order result is the same for bosons and fermions. This is expected, as it corresponds to the single-particle
contributions, and therefore the statistics plays no role.

\begin{figure}[h!]
\includegraphics[width=1.0\columnwidth]{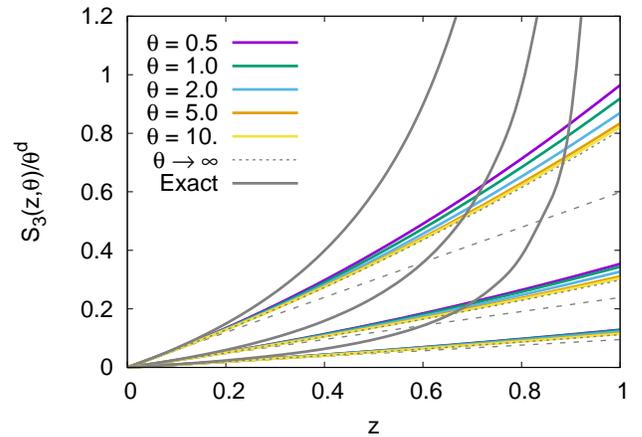}
\caption{\label{Fig:S3FugacityBosons}(color online) 
Third R\'enyi entropy $S_3^{}$ of noninteracting bosons as a function of the fugacity and for several values
of the dimensionless region size $\theta$ and for dimensions 1, 2 and 3 (top to bottom groups). The dashed lines show,
for each dimension, the leading order results, which are proportional to $\theta^d$.
}
\end{figure}

In Fig.~\ref{Fig:S2FugacityBosons} we show our results for bosons for the $n=2$ case. 
As in the fermionic case, there is no dependence on $\theta$ for $n=2$.
In Fig.~\ref{Fig:S3FugacityBosons}, we show the $n=3$ case for several values of the dimensionless region size $\theta$ and 
for dimensions 1, 2 and 3. In contrast to the $n=2$ case, $n>2$ does have non-trivial volume effects at second order in $z$.

%

\section{Summary and conclusions}

We have performed a semi-analytic calculation to all orders in the fugacity $z=e^{\beta\mu}$ of the R\'enyi entanglement 
entropy of noninteracting quantum gases in arbitrary spatial dimensions. 
We provide an analytic expression in the continuum, high-temperature limit $L\to\infty$,
$\lambda_T^{}/L\to0$, at fixed $\lambda_T^{}/L_A^{}$. 
In addition, we have calculated the leading and next-to-leading order terms in the virial expansion.

As expected,
we found volume scaling for our exact results as well as at both orders in the virial expansion.  At leading order (where the statistics is irrelevant) 
the result is the same for bosons and fermions. For fermions, the two-particle, i.e. $z^2$, contributions
decrease the entanglement entropy relative to the leading-order result, while the opposite happens
for bosons; this appears to be true for all $n$.

In all cases, we find that the convergence properties of the virial expansion, as estimated from considering
the two orders computed here, improve when going to higher dimensions. Indeed, in the scale shown in
our plots the results for $d=1$ are much less promising in this regard than $d=3$.

With the rapid development in the area of Monte Carlo calculations of the entanglement properties
of many-body systems, having non-perturbative results available is of great value as a benchmark. 
In this work we have aimed to provide such results.

\acknowledgements

The authors are grateful to L. Rammelm\"uller for thought-provoking discussions and proofreading.  This material is based upon work supported by the National Science Foundation under
Grants
No. PHY1306520 (Nuclear Theory program)
and
No. PHY1452635 (Computational Physics program). 



\end{document}